\documentclass[11pt,a4paper]{article}
\usepackage{delarray,amsfonts,amsmath,amssymb,graphicx}
\DeclareGraphicsExtensions{ps,eps,ps.gz}

\newcommand{\norm}[1]{\left\lVert#1\right\rVert}
\newcommand{\abs}[1]{\lvert#1\rvert}
\def\aa{{\cal A}}
\def\hh{{\cal H}}
\def\rr{{\mathbb{R}}}
\def\cc{{\mathbb C}}
\def\pp{{\cal P}}
\def\re{\text{Re}}
\def\im{\text{Im}}

\makeatletter
\def\@citex[#1]#2{\if@filesw\immediate\write\@auxout
        {\string\citation{#2}}\fi
\def\@citea{}\@cite{\@for\@citeb:=#2\do
        {\@citea\def\@citea{,}\@ifundefined
        {b@\@citeb}{{\bf ?}\@warning
        {Citation `\@citeb' on page \thepage \space undefined}}
        {\csname b@\@citeb\endcsname}}}{#1}}
\newif\if@cghi
\def\cite{\@cghitrue\@ifnextchar [{\@tempswatrue
        \@citex}{\@tempswafalse\@citex[]}}
\def\citelow{\@cghifalse\@ifnextchar
[{\@tempswatrue\@citex}{\@tempswafalse\@citex[]}}
\def\@cite#1#2{{\if@cghi\unskip$\null^{#1}$\else #1\fi\if@tempswa\typeout
        {warning: optional citation argument ignored: `#2'} \fi}}

\begin{document}
\title{Smoother than a circle\\ {\small{or}} \\ How Noncommutative Geometry provides
the torus with an egocentred metric}
\author{Pierre Martinetti,\\ Centre de physique th\'eorique, CNRS Luminy,
Marseille\\{\it pierre.martinetti@laposte.net}}  \maketitle
\date\begin{abstract}
 {\it This is a non-technical version of [\citelow{cc}] written as
proceedings of the international conference on "Differential
Geometry and its Applications" held in Deva, Romania, on October
2005. Published by Cluj university press.} \newline We give an
overview on the metric aspect of noncommutative geometry, especially
the metric interpretation of gauge fields via the process of {\it
fluctuation of the metric}. \linebreak Connes' distance formula
associates to a gauge field on a bundle $P$ equipped with a
connection $H$ a metric. When the holonomy is trivial, this distance
coincides with the horizontal distance defined by the connection.
When the holonomy is non-trivial, the noncommutative distance has
rather surprising properties. Specifically, we exhibit an elementary
example on a $2$-torus in which the noncommutative metric $d$ is
somehow more interesting than the horizontal one since $d$ preserves
the $S^1$-structure of the fiber and also guarantees the smoothness
of the length function at the cut-locus. In this sense the fiber
appears as an object "smoother than a circle". As a consequence,
from the intrinsic metric point of view developed here, any observer
whatever his position on the fiber can equally pretend to be "the
center of the world".
\end{abstract}

\begin{figure}[h]
\begin{center}
\mbox{\rotatebox{0}{\scalebox{.75}{\includegraphics{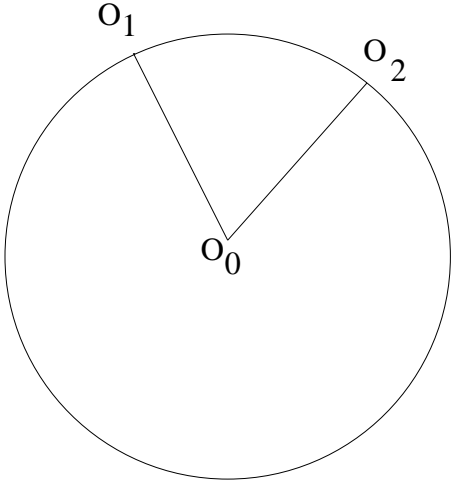}}}}
\caption{\label{cercentre}The expression "center of $S^1$" is
meaningful only when the  $1$-dimensional object is embedded within
an $d>1$-dimensional space.}\end{center}
\end{figure}

\begin{figure}[h]
\begin{center}
\mbox{\rotatebox{0}{\scalebox{.75}{\includegraphics{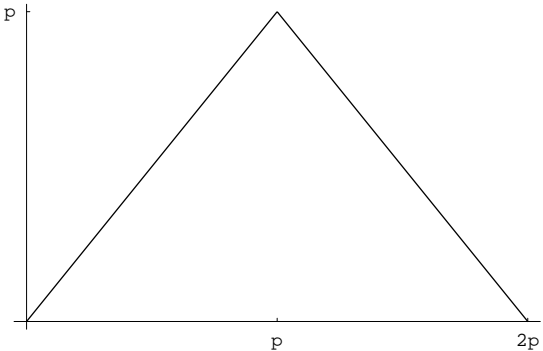}}}}
\caption{\label{deuclid}The Euclidean distance on a circle of radius
$1$.}\end{center}
\end{figure}

\begin{figure}[h]
\begin{center}
\mbox{\rotatebox{0}{\scalebox{.75}{\includegraphics{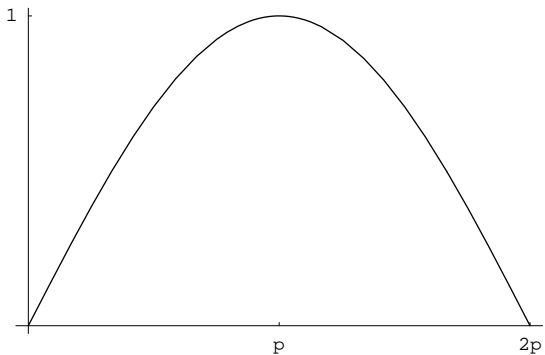}}}}
\caption{\label{dnc}A possible metric for an object smoother than
a circle.}
\end{center}
\end{figure}

Let us begin with a toy-cosmology tale. Consider the simplest
cosmological model, namely a $1$-dimensional universe homeomorphic
to a circle $S^1$. The main question for cosmologists is to
determine the shape of this universe. Does it look like an Euclidean
circle, an egg, or some formless potato ? Consider one cosmologist,
say $O_1$, located somewhere in this universe and assume his main
information about the surrounding world consists in the measurement
of the distance that separates him from any other points. For
simplicity, and because he believes he is the center of the world,
$O_1$ has chosen a parameterization $\phi_1\in[0,2\pi[$ of $S^1$
such that $\phi_1 = 0$ corresponds to its own position. Hence all
that he knows about the world is encoded within the function
$$d_1(\phi_1) = \text{ distance between $0$ and $\phi_1$.}
$$
The same is true for another observer $O_2$ located at $\phi_1\neq
0$. Also believing he is the center of the world, $O_2$ has chosen a
parameterization $\phi_2$ such that zero corresponds to his own
position. His knowledge is entirely encoded within a function
$d_2(\phi_2)$. Now assume that our observers respectively find
$$d_1(\phi_1) = \min(\phi_1, 2\pi-\phi_1),\quad
d_2(\phi_2) = \min(\phi_2, 2\pi-\phi_2).$$ They will agree that the
universe is indeed a Euclidean circle. But they won't agree on who
is actually the center of the world. Both of them can equally
pretend to be at some very particular point $\phi_i=0$. Then comes a
third cosmologist $O_0$, a theoretician, who explains that the
quarrel has no physical meaning and is only a matter of
parameterization. Everybody knows, $O_0$ says, that the notion of
center has no meaning {\it on} a Euclidean circle. All points are on
the same footing with respect to each other and the only way to
recover a notion of "center" comes from extra-dimensions, for
instance by embedding the $1$-dimensional universe into a
$2$-dimensional Euclidean plane (figure \ref{cercentre}).
Also, one of the $O_i$'s may argue that the universe they see does
not look very smooth since neither of the $d_i$ functions is smooth
(see figure \ref{deuclid}).
In fact, from an intrinsic metric point of view, a Euclidean circle
is a collection of discontinuities (any point is a cut-locus for a
given observer). As physics is keen on smoothness, our cosmologists
may be worried and wonder what metric, if any, could make their
$S^1$-universe smoother. It would be something close to the
Euclidean metric but that avoids the discontinuity of the derivative
at $\phi_i=\pi$ (figure \ref{dnc}). For instance would it be
possible to measure
\begin{equation}
\label{question1}
 d_i(\phi_i) = \sin \frac{\phi_i}{2}
 \end{equation}
and, in this case, what would be the shape of such a
smoother-than-a-circle object ?

 Let us now forget our cosmology tale for a while and ask a more serious - and apparently
 unrelated - question.
In noncommutative geometry\cite{connes} the metric information is
encoded within the Dirac operator. Specifically, via Connes'
distance formula (see eq.\ref{distance} below) one is able to
recover from purely algebraic data the geodesic distance on a
riemannian compact smooth spin manifold $M$,
$$-i\gamma^\mu\partial_\mu
\Longleftrightarrow  \text{Riemannian geodesic distance.}$$ Physics
not only deals with spin manifolds but also with gauge theories,
that is to say bundles $P$ equipped with a connection $H$ (and an
associated $1$-form $A_\mu$). Therefore it is quite natural to
wonder what distance $d$ is encoded within the covariant Dirac
operator,
\begin{equation}\label{question2}
-i\gamma^\mu(\partial_\mu + A_\mu) \Longleftrightarrow \, ?
\end{equation}
It turns out that for a very simple example of covariant Dirac
operator, the distance $d$ is precisely the one expected by our
cosmologists in (\ref{question1}). This example is treated in detail
in [\citelow{cc}]. We give here a non-technical account of this
result.
\newline

Before entering into the details let us recall that gauge fields
already have a well known metric interpretation in terms of
horizontal distance $d_H$. This distance, also called {\it
Carnot-Carath\'eodory} or sub-Riemannian distance\cite{montgomery},
is by definition the length of the shortest path whose tangent
vector is everywhere horizontal with respect to the connection $H$
{\footnote{A connection $H$ is the splitting of the tangent space
$TP$ into a horizontal subspace (the kernel of the connection
$1$-form) and a vertical subspace,
$$T_pP = V_pP\oplus H_pP.$$}}(see figure
\ref{dcc}),
\begin{equation}
\label{dcc} d_H(p,q) = \underset{\dot{c}(t)\in H_{\!c(t)}\!
P}{\text{Inf}}\; \int_0^1 \norm{\dot{c}(t)} dt
\end{equation}
where $c(0)=p$, $c(1)=q$ is a smooth curve in $P$. In
[\citelow{gravity}] Connes has pointed out the link between $d_H$
and $d$. In [\citelow{cc}] we have examined this link in detail,
showing the importance of the holonomy of the connection on this
matter. We also worked out an example in which the two distances are
not equal. This is this example, related to the toy-cosmology metric
(\ref{question1}), that we study in detail below.
\begin{figure}[h]
\begin{center}
\mbox{\rotatebox{0}{\scalebox{.9}{\includegraphics{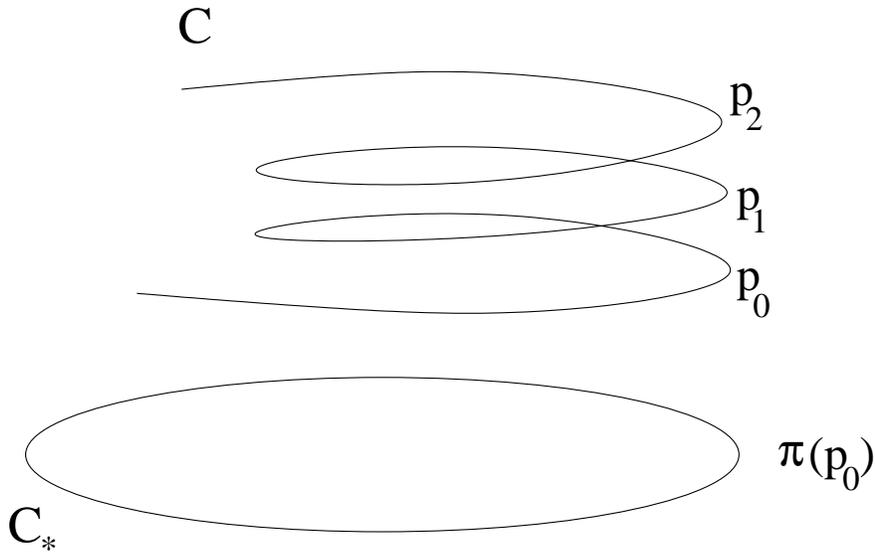}}}}
\end{center}
\caption{With shortest horizontal path a helix of radius $1$,
$d_H(p_0, p_2) = 4\pi$.}
\end{figure}

\section{The distance formula in noncommutative geometry}

Noncommutative geometry aims at understanding the geometry of a
space whose algebra of functions (defined on it) is noncommutative.
Such objects are well described in terms of spectral triples
$(\aa,\hh,D)$, where $\aa$ is an associative $*$-algebra
(commutative or not) represented by $\pi$ on a Hilbert space $\hh$
and $D$ is an operator on $\hh$. Together with a chirality $\Gamma$
and a real structure $J$ also acting on $\hh$, these three elements
satisfy a set of 7 conditions\cite{gravity} providing the necessary
and sufficient conditions for

1) an axiomatic definition of Riemannian spin geometry in terms of
commutative algebra,

2) its natural extension to the noncommutative framework.

 \noindent Explicitly, given a spin manifold $M$ one checks that
$$(C^{\infty}\!(M), L_2(M,S), -i\gamma^\mu\partial_\mu)$$
is a spectral triple, with $L_2(M,S)$ the space of square integrable
spinors on $M$. Conversely star\-ting from a spectral triple $(\aa,
\hh, D)$ with $\aa$ the algebra of smooth functions over a compact,
Riemannian manifold $N$, one obtains that $N$ is indeed a spin
manifold with corresponding Dirac operator $D$ (modulo a torsion
term). Moreover the geodesic distance corresponding to the
Riemannian structure of $N$ is given by{\footnote{To get familiar
with this formula, one can study the example $N=\rr$,
$D=\frac{d}{dx}$, $\hh = L_2(\rr)$. Then $\norm{[D,f]}= \norm{f'} =
\underset{z\in\rr}{\sup}\{\abs{f'(z)}\}$ so that
$(\ref{distance})\leq \abs{x-y} =d_{geo}(x,y).$ This upper bound is
attained by the function $z\mapsto z$.}}
\begin{equation}
\label{distance} d_{geo}(x,y) =  \underset{f\in
C^\infty(N)}{\sup}\{ \abs{f(x) - f(y)}\, \slash \norm{[D,f]}\leq
1\}.
\end{equation}
Extension to the noncommutative framework is obtained by dropping
the commutativity of $\aa$. "Points" are recovered as pure
states{\footnote{State: positive linear mapping $\tau$ from $\aa$ to
$\cc$. Pure state $\omega$: state that does not decompose as  a
convex combination of other states, $\omega\neq \lambda \tau_1 +
(1-\lambda) \tau_2.$}} $\pp(\aa)$ of $\aa$, in analogy with the
commutative case where, by Gelfand duality, $\pp(C^\infty(M)) \simeq
M$ with explicit homeomorphism
\begin{equation}
\label{gelfand} x\mapsto\omega_x \text{ such that } \omega_x(f)
\doteq f(x)
\end{equation}
for any $f\in C^\infty(M)$. Formula (\ref{distance}) rewritten
as\cite{metrique}
\begin{equation}
\label{distance2} d(\omega_1,\omega_2) =\underset{a\in
\aa}{\sup}\{ \abs{\omega_1(a) - \omega_2(a)}\, \slash
\norm{[D,a]}\leq 1\}\end{equation}
 defines a distance $d$ between states which

-makes sense whether $\aa$ is commutative or not,

-is consistent with the classical case, $d=d_{geo}$, when
$\aa=C^\infty(M)$,

-does not involve notions ill-defined in a quantum context such as
paths between points.

In [\citelow{finite}] we computed $d$ in an $n$-point space
($\aa=\cc^n$) as well as for other finite dimensional algebras. For
instance $\aa=M_2(\cc)$ yields a metric on the
$2$-sphere{\footnote{$\pp(M_2(\cc))=\cc P^{1}$ is in one to one
correspondence with $S^2$; see eq.(\ref{hopf}) below.}}. In fact
finite dimensional spectral triples
 are particularly interesting in products of
geometries. Namely given a spin manifold $M$ and a spectral triple
$(\aa_I, \hh_I, D_I)$, one defines
\begin{eqnarray}
 \aa= C^\infty(M)\otimes\aa_I,
 \hh= L_2(M,S)\otimes\hh_I,
\label{product} D= -i\gamma^\mu\partial_\mu \otimes \mathbb{I}_I +
\gamma^5\otimes D_I
\end{eqnarray}
which again is a spectral triple. $\pp(\aa)$ is the set of couples
$$(\omega_x\in\pp(C^\infty(M)),\, \omega\in\pp(\aa_I)),$$ so that for a
finite dimensional $\aa_I$ the spectral triple (\ref{product})
describes a geometry which is the product of a discrete space
$\pp(\aa_I)$ by a continuous one $\pp(M)$. For instance $\aa_I =
\cc^2$ yields a two-sheeted model, two copies of $M$ indexed by the
pure states of $\cc^2$. On each copy the distance (\ref{distance2})
coincides with the geodesic distance of $M$ while it remains finite
between the sheets, although there is no "path" between them. Note
that some metric aspects of sums, rather than products, of spectral
triples have been studied very recently\cite{christensen}.

\section{Fluctuation of the metric}

Inspired by the commutative case, $Diff(M) = Aut(C^\infty(M))$,
one describes the symmetries of a noncommutative geometry in terms
of automorphisms of $\aa$. The group $Aut(\aa)$ naturally splits
into inner automorphisms,
$$In(\aa): \alpha_u(a) = uau^*$$
given by unitary elements ($uu^*=\mathbb{I}$) of $\aa$, and outer
automorphisms
$$Out(\aa)=Aut(\aa)/In(\aa).$$
The latter have a nice interpretation in quantum field theory as
flow of time\cite{outemps}. The former are characteristic of the
noncommutative case (otherwise $In(\aa)$ is trivial). Remarkably,
the action of $In(\aa)$ on a geometry $(\aa, \hh, D)$ via the
replacement of the representation $\pi$ by $\pi\circ\alpha_u$ is
equivalent to replacing $D$ by
\begin{equation}
\label{fluctd} D_A \doteq D + A + JAJ^{-1}\end{equation} where
$A\doteq u[D,u^*]$.
 This appears as a particular instance of the
so-called {\it fluctuations of the metric}, defined in a more
general manner by taking
\begin{equation}
\label{potvect} A = \underset{i}{\Sigma}\, a_i[D,b_i]\quad a_i,
b_i \in \aa .
\end{equation}

 In the case of the product geometry (\ref{product}), explicit
 computations\cite{ktvh} yield
 $$A = H -i\gamma^\mu A_\mu$$
  where $H$ is a scalar field on $M$
with value in $\aa_I$ (the Higgs field in the standard
model\cite{spectral}) and $A_\mu$ is a $1$-form field with value
in $Lie(U(\aa_I))$, that is to say a gauge field. Therefore via
the fluctuations of the metric both the Higgs field and gauge
fields acquire a metric interpretation. In [\citelow{kk}] we
focused on the Higgs field only, assuming $A_\mu = 0$. In
[\citelow{gravity}] Connes considers the example $H=0$ with an
internal geometry $\aa_I = M_n(\cc)$. The vanishing of $H$ is
obtained by taking $D_I=0$ so that the fluctuated Dirac operator
$$D_A = -i\gamma^\mu(\partial_\mu + A_\mu)$$
is nothing but{\footnote{Note the slight abuse of notation: we
canceled out the $JAJ^{-1}$ term in (\ref{fluctd}) since it
commutes with the representation and so does not play any role in
the computation of the distance.}} the usual covariant Dirac
operator on the $U(n)$ trivial bundle
\begin{equation}
\label{P} P =M\times \cc P^{n-1}
\end{equation}
with connection $1$-form $A_\mu$. The latest defines both a
Carnot-Carath\'eodory distance $d_H$ via (\ref{dcc}) and a
noncommutative distance $d$ via (\ref{distance2}) (with $D_A$
instead of $D$). It is not difficult to show that whatever $M$ and
$n$, $d\leq d_H$. Also $d = d_H$ as soon as the holonomy of the
connection is trivial. However when the holonomy is non trivial $d$
does not necessarily equals $d_H$. We examine below a simple
example, the $2$-torus $S^1\times S^1$, in which the two distances
have strongly different characteristics: while $d_H$ forgets about
the bundle structure of $P$, $d$ deforms the Euclidean metric on the
fiber $S^1$ in a rather intriguing way, providing it with the
toy-cosmology metric (\ref{question1}).

\section{An egualitarian and vanity-preserving metric}

Take $M=S^1$ and $\aa_I= M_2(\cc)$, that is to say
$$\aa = C^\infty(M)\otimes M_2(\cc) \approx C^\infty(M,M_2(\cc)). $$
$\pp(\aa)$ is the trivial bundle on the circle with fiber $\cc
P^1$, mapped to the $2$-sphere via the Hopf map \begin{equation}
\label{hopf} \xi=
\begin{array}(c) \xi_1\\\xi_2\end{array}\in \cc P^1
\mapsto \begin{array}(c) x_\xi = 2\re(\xi_1\bar{\xi_2})\\
y_\xi = 2\im(\xi_1\bar{\xi_2})\\ z_\xi = \abs{\xi_1}^2
-\abs{\xi_2}^2\end{array} \in S^2. \end{equation}
 Let us fix a trivialization on $P$ and write $\xi_x$ for the point in the fiber over
 $x$ corresponding to $\xi\in \cc P^1$. Take
$$A_\mu = \left(\begin{array}{cc} 0&0\\0&\theta\end{array}\right)
$$
with $\theta\in]0,1[$. Then for any $\xi_x, \zeta_y \in P$ one
easily computes that
$$d(\xi_x,\zeta_y)=+\infty$$
if and only if $z_\xi\neq z_\zeta$. Hence the set of pure states at
finite noncommutative distance from $\xi_x$ is a two-torus
$$T_\xi = M\times S^1.$$
\begin{figure}[h*]
\begin{center}
\mbox{\rotatebox{0}{\scalebox{.5}{\includegraphics{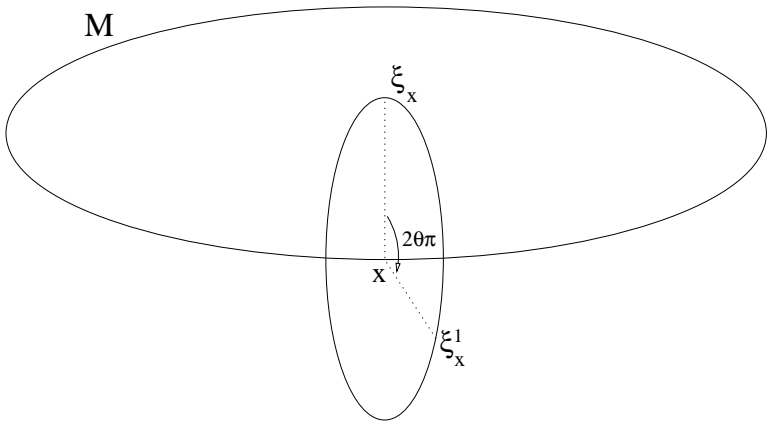}}}}
\caption{\label{figuretore} The torus $T_\xi$.}
\end{center}
\end{figure}

\begin{figure}[h]
\begin{center}
\mbox{\rotatebox{0}{\scalebox{.75}{\includegraphics{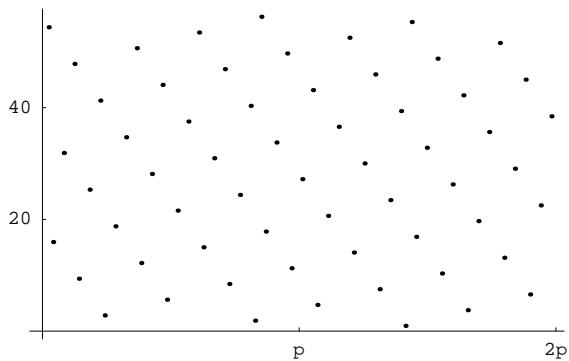}}}}
\caption{\label{figdh}The horizontal distance $d_H$ on the fiber.
Vertical unit is $C$.}
\end{center}
\end{figure}

\noindent Let us parameterize the $S^1$ fiber by $\phi\in[0,2\pi[$
such that
\begin{equation}
\label{paramun} 0= \xi_x= \xi_x^0
\end{equation}
 and define
$\xi_x^k \doteq 2k\theta\pi$ as the end point of the horizontal
lift starting at $\xi_x^{k-1}$ of the basis $S^1$ (see figure
\ref{figuretore}). Assuming the basis has radius $1$, the
horizontal distance is
$$d_H(\xi_x,\xi_x^k)= 2k\pi.$$ In case $\theta$ is irrational, any
neighborhood of $0$ contains some $\phi_k \doteq 2k\theta\pi
\text{ mod }[2\pi]$ with $k$ arbitrarily large. Hence, as plotted
in figure \ref{figdh}, $d_H$ "destroys" the $S^1$ structure of the
fiber.
On the contrary a rather long calculation carried out in
[\citelow{cc}] shows that
\begin{equation}
\label{ccresult} d(0, \phi)= C \sin \frac{\phi}{2}
\end{equation}
for any $\phi\in [0,2\pi[$, with
$$C=\frac{4\pi\abs{\bar{\xi_1}\xi_2}}{\abs{\sin \theta\pi}}$$ a
constant. This is nothing but the metric expected in
(\ref{question1}), which is smooth at the cut locus $\phi = \pi$.
Hence from the metric point of view the fiber over $x$ equipped
with the noncommutative distance $d$ is smoother than a circle.

\begin{figure}[h]
\begin{center}
\mbox{\rotatebox{0}{\scalebox{1.3}{\includegraphics{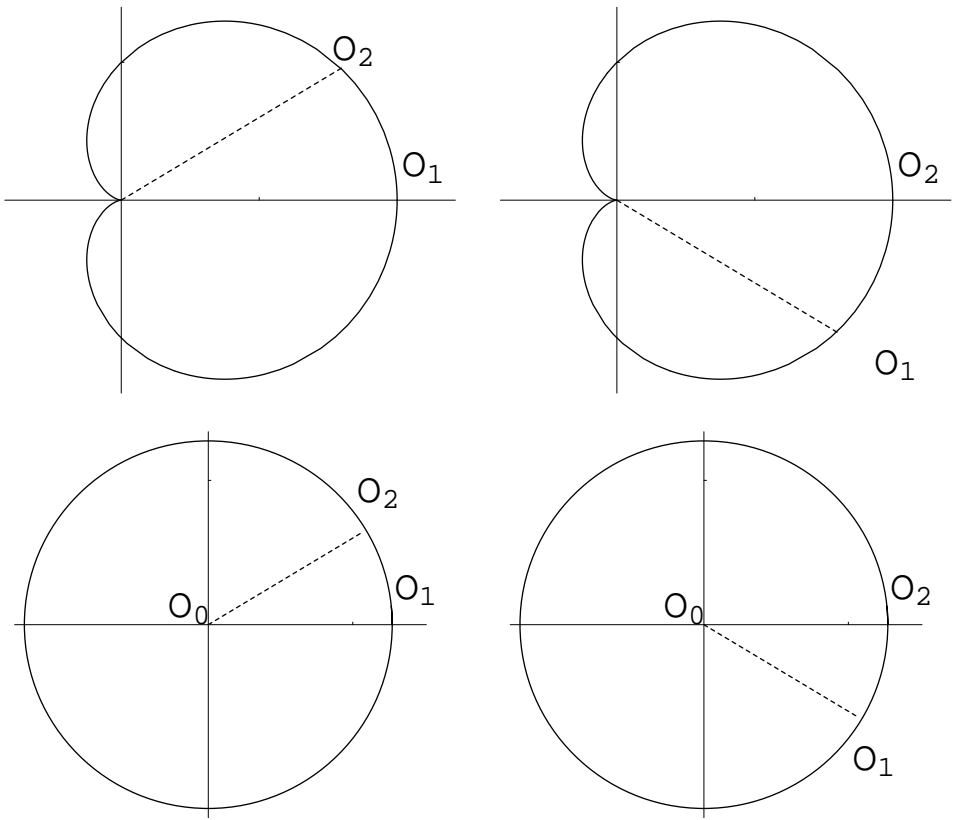}}}}
 \caption{\label{according}On the left, the world according to
$O_1$, who occupies the position $\phi_1=0$. On the right, the world
according to $O_2$ (located at $\phi_2=0$). At top the world is a
cardioid, $O_1$'s and $O_2$'s visions are not compatible with an
embedding of the $S^1$-fiber into a 2-dimensional Riemannian space
(the two cardioids do not coincide). At bottom the world is a
Euclidean circle. From the intrinsic points of view of the $O_i$'s
as well as for an $O_0$ outside observer embedded into the Euclidean
plane, the two objects coincide up to a rotation.}
\end{center}
\end{figure}

There still remains the question asked by our cosmologists: what
does $S^1$ equipped with (\ref{ccresult}) look like?
\begin{figure}[h]
\begin{center}
\mbox{\rotatebox{0}{\scalebox{.55}{\includegraphics{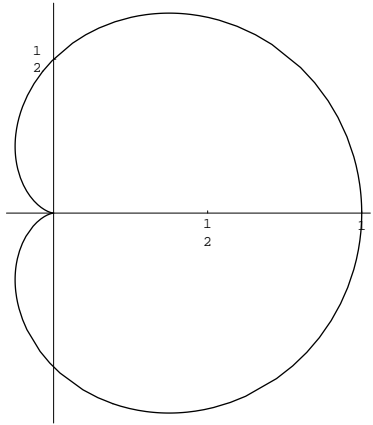}}}}
\caption{\label{cardiofig} A cardioid.}
\end{center}
\end{figure}In polar coordinates
$d(0,\phi)$ is the Euclidean length on the cardioid $\frac{C}{4}(1 +
\cos\phi)$ (see figure \ref{cardiofig}). So one could be tempted to
believe that the fiber is indeed a cardioid. One has to be careful
with this interpretation. The noncommutative distance $d$ is
invariant under translation on the fiber: the identification of
$\xi_x$ with $\phi=0$ in (\ref{paramun}) is arbitrary; identifying
$0$ with $\zeta_x$ with $\zeta\neq\xi$ and $z_\xi = z_\zeta$ would
lead to a similar result $d(0,\phi)= C\sin \frac{\phi}{2}$. On the
contrary the Euclidean distance on the cardioid is not invariant
under translation. Therefore, assuming that our observers $O_1, O_2$
are respectively located at $\xi_x$ and $\zeta_x$, each of them
measures $d_i(\phi_i) = d(0,\phi_i)$, $i=1,2$. Both agree that the
fiber they are lying on is a cardioid and both pretend to be
localized at this particular point opposite to the "cusp" of the
cardioid (see figure \ref{according}).
Both are equally right and the quarrel is not just a matter of
parameterization as on the Euclidean circle. On a cardioid all
points are not on the same footing, the cusp and the point opposite
to the cusp in particular have unique properties (they are their own
image under the axial symmetry $\sigma$ that leaves the cardioid
globally invariant). The very particular position of $O_1$ according
to its own point of view (he is his own image by $\sigma$) seems in
contradiction with $O_2$'s point of view (for whom $O_1$ is not
$\sigma$-invariant). Moreover, extra-dimensions are of no help:
measuring (\ref{question1}) is not compatible with some $O_0$'s
"outside" riemannian point of view. In fact the contradiction only
comes from the implicit assumption that the fiber is a riemannian
manifold. What (\ref{ccresult}) shows is that the fiber equipped
with $d$ is not a riemannian manifold. It is a geometry in which
everyone can on the same footing pretend to be the center of the
world and in which, at the same time, the notion of center (i.e. of
points with particular symmetry properties) still has a meaning. In
brief, whereas the price to pay for equality on the Euclidean circle
is either to renounce the notion of center or to recover it from an
extra-dimension (both options are hard for $O_i$'s own vanity), the
noncommutative distance on the fiber of $T_\xi$ is both egualitarian
and vanity-preserving.

\section{Conclusion}

As a conclusion let us mention several groups of questions. First,
there is still a lot to do to clarify the physical meaning of this
"smoothness from an intrinsic metric point of view". In particular
it would be interesting to check whether the same properties can be
observed with a fiber of higher dimension, like the $2$ or the $3$-
sphere. Similarly one should deal with other base-manifolds than
$S^1$. As explained in [\citelow{cc}], the link between the holonomy
of the connection and the possibility for the noncommutative
distance to equal the Carnot-Carath\'eodory one yields a nice
question for sub-Riemannian geometers\cite{montgomery}: given a
minimal horizontal curve, is it possible to deform it keeping its
length fixed and reducing the number of times its projection on $M$
self-intersects ? The answer is obviously "no" for $M$ of dimension
$1$ but seems unknown for dimension greater than 3.

Another open question is to compute the metric when both the scalar
part $H$ of the fluctuation and the gauge part $A_\mu$ are different
from $0$. Finally let us underline that [\citelow{cc}] was intended
to be a preliminary step towards the study of the metric aspect of
the noncommutative torus\cite{rieffel}. The situation there should
be quite similar, except that the pure state space is then a twisted
bundle.
\newline

\noindent{\bf Acknowledgments} Thanks to P. Almeida and Y. Fr\'egier
for discussions on the interpretation of these results. Thanks to A.
Greenspoon for a careful reading and correcting of the manuscript.

\end{document}